\documentclass[12pt]{article}
\usepackage{epsf}
\usepackage[dvips]{color}
\usepackage{exscale}
\usepackage[dvips]{graphicx}
\usepackage{amsmath}
\usepackage{amssymb}
\usepackage{latexsym}

\begin{document}
\setlength{\baselineskip}{18pt}
\begin{titlepage}
\begin{flushright}
EPHOU-13-001
\end{flushright}

\vspace*{1.2cm}
\begin{center}
{\Large\bf Separate seesaw and its applications to\\dark matter and baryogenesis}
\end{center}
\lineskip .75em
\vskip 1.5cm

\begin{center}
{\large Ryo Takahashi}

\vspace{1cm}

{\it Department of Physics, Faculty of Science, Hokkaido University, Sapporo 
060-0810, Japan}

\vspace*{10mm}
{\bf Abstract}\\[5mm]
{\parbox{13cm}{
We propose a new seesaw model in an extra-dimensional setup where only 
right-handed neutrinos are bulk fields. In the model, the localizations of an 
extra-dimensional wave function and brane Majorana mass of the right-handed 
neutrinos can be different among each generation of the right-handed neutrinos. 
The setup can lead to different suppression factor dependences of effective  
right-handed neutrino masses and neutrino Yukawa couplings for each generation.
 It is shown that the resultant mass spectra of the right-handed neutrinos and 
neutrino Yukawa couplings are favored in models of neutrino dark matter with baryogenesis.
}}
\end{center}
\end{titlepage}

\tableofcontents

\section{Introduction}
The results from neutrino experiments have established the phenomenon of 
neutrino oscillations. The experimental data can be nicely understood by three 
flavor mixings of active neutrinos and described by three mixing angles, one 
CP-phase, and two mass-squared differences of the active neutrinos. The non-zero
 masses of the active neutrinos require a physics beyond the standard model 
(SM). In addition, one has to explain the smallness of the active neutrino mass 
scales. One famous explanation for realizing the small active neutrino masses 
is the seesaw mechanism~\cite{seesaw}. There are several types of seesaw 
mechanism. In the canonical type-I seesaw mechanism, heavy right-handed 
(sterile) neutrinos are introduced into the SM as Majorana particles. After 
decoupling the heavy right-handed neutrinos, the small active neutrino masses 
can be realized. Furthermore, it is well known that decays of such heavy 
Majorana (right-handed) neutrinos can also generate the baryon asymmetry of the 
universe (BAU), so-called leptogenesis~\cite{Fukugita:1986hr}.

In addition to leptogenesis, the right-handed (sterile) neutrinos can also 
play a cosmological role as a dark matter (DM)~\cite{Zwicky:1933gu} 
candidate. If the mass of the sterile neutrino is in the keV range, the sterile 
neutrino can serve as DM~\cite{Dodelson:1993je,Shi:1998km} (see 
also~\cite{Dolgov:2000ew,Abazajian:2001nj}) and such a sterile neutrino DM
 model is constrained by cosmological and astrophysical 
observations~\cite{Dolgov:2000ew,Mapelli:2006ej,RiemerSorensen:2006pi,Kusenko:2009up}. There are also 
some investigations of astrophysical 
phenomena~\cite{Kusenko:2009up,Kusenko:1997sp} and related considerations for the keV sterile neutrino DM (e.g., see~\cite{Barry:2011fp,Merle:2011yv,Kusenko:2010ik,Adulpravitchai:2011rq,BRZ}). One interesting discussion about the keV sterile neutrino DM gives simultaneous explanations for problems in particle physics and cosmology such as the smallness of the active neutrino masses, the origin of BAU, and LSND/MiniBooNE anomaly~\cite{Aguilar:2001ty} in addition to the DM~\cite{Asaka:2005an,Kusenko:2010ik,Adulpravitchai:2011rq,BRZ}. However, such models for  simultaneous explanations of several problems require strongly hierarchical mass spectra of the sterile neutrinos and the additional suppression of left-right mixing of the keV sterile neutrino in the context of the seesaw mechanism. A split seesaw~\cite{Kusenko:2010ik}\footnote{See also~\cite{Adulpravitchai:2011rq} for a consideration of $A_4$ flavor models in the split seesaw mechanism.} and the Froggatt-Nielsen mechanism~\cite{Froggatt:1978nt} can naturally realize strongly hierarchical mass spectra of the right-handed neutrinos~\cite{Merle:2011yv} but do not give the additional suppression of the left-right mixing without taking the corresponding neutrino Yukawa coupling to be small enough to satisfy the cosmological bound. In the split seesaw mechanism, an essential point to obtain hierarchical mass spectra comes from the properties of the wave function in an extra dimension.

Extra-dimensional theory is one of the fascinating approaches to afford a compelling solution to the gauge hierarchy problem~\cite{ArkaniHamed:1998rs}. In fact, the discovery of evidence of extra dimension(s) in addition to the Higgs particle and supersymmetry is one of the missions of the Large Hadron Collider experiment. Further, extra-dimensional theories can also give rich phenomenologies such as alternatives to the electroweak (EW) symmetry breaking mechanisms in the SM (e.g., see~\cite{Fairlie:1979at,Haba:2009pb}), DM candidates~\cite{Antoniadis:1990ew,Holthausen:2009qj}, deviations of coupling constants from standard ones~\cite{Haba:2009pb,Hosotani:2008tx}, and realizations of fermion mass hierarchies~\cite{Holthausen:2009qj,Dienes:1998vh} and the smallness of the active neutrino mass~\cite{Dienes:1998sb}, etc. In this work, we will propose a new model with right-handed neutrinos in an extra-dimensional space and apply it to models of keV sterile neutrino DM with the baryogenesis mechanism. Then, it will be shown that the new seesaw can be favored in the DM models with respect to a realization of the additional suppression of left-right mixing.

The paper is organized as follows: In Sect. 2, brief reviews of the canonical type-I seesaw and the split seesaw mechanisms are given. In Sect. 3, we discuss all possible localizations of the right-handed neutrinos and Majorana masses in an extra-dimensional space-time, and show the results for relevant parameters to the neutrino masses. In Sect. 4, a new seesaw model from the extra dimension is proposed. In Sect. 5, we apply the new seesaw to models of the neutrino DM with baryogenesis mechanism. Section 6 is devoted to the summary.

\section{Seesaw mechanisms}

\subsection{Type-I seesaw mechanism}

The relevant Lagrangian for the type-I seesaw mechanism~\cite{seesaw} reads
 \begin{equation}
  {\cal L}=i\overline{\nu_{R_i}}\gamma^\mu\partial_\mu\nu_{R_i}
           -\left((y_\nu)_{i\alpha}\overline{\nu_{R_i}}L_\alpha\phi 
                  +\frac{(M_R)_{ij}}{2}\overline{\nu_{R_i}^c}\nu_{R_j}
                  +{\rm h.c.}\right),
 \end{equation}
where $\nu_{R_i}$ ($i$=1,2,3) are the right-handed neutrinos, $L_{\alpha}$ 
($\alpha=e,\mu,\tau$) are left-handed lepton doublets, $\phi$ is the SM Higgs, 
$y_\nu$ is the neutrino Yukawa coupling matrix, $M_R$ is the Majorana mass matrix of
 the right-handed neutrinos, and $\mu=0,1,2,3$, respectively. Here we have 
introduced three generations of right-handed neutrinos to the SM. After 
integrating out the right-handed neutrinos, the mass matrix of light (active) 
neutrinos is given by
 \begin{equation}
  M_{\nu}\simeq y_\nu^TM_R^{-1}y_\nu v^2,
 \end{equation}
where $v$ is the vacuum expectation value (VEV) of the Higgs. This type-I seesaw 
mechanism with heavy Majorana masses of the right-handed neutrinos can lead to small active neutrino masses without tiny neutrino Yukawa couplings. 

\subsection{Split seesaw mechanism}

Next, let us show a seesaw model in an extra dimension. This seesaw model is 
known as the split seesaw mechanism~\cite{Kusenko:2010ik}. The split seesaw has been
 considered in a flat 5D space-time compactified on an orbifold $S^1/Z_2$ with 
the 5D coordinate, $y\equiv x^5$. A fundamental region 
of the orbifold is given by $y\in[0,\ell]$. The orbifold gives two fixed points and one can consider 
branes at the fixed points. One of the branes $(y=0)$ is an SM brane where the 
SM particles reside, while the other $(y=\ell)$ is a hidden brane.

Then three generations of Dirac spinor, 
$\Psi_i(x,y)=(\chi_i(x,y),\bar{\psi}_i(x,y))^T$, are introduced on the bulk 
with bulk masses $m_i$ in this flat 5D setup. A fundamental action for the 5D 
Dirac spinor is 
 \begin{equation}
  S=\int d^4x\int_0^\ell dyM(i\overline{\Psi_i}\Gamma^A\partial_A\Psi_i
                             +m_i\overline{\Psi_i}\Psi_i),\qquad A=0,1,2,3,5,
  \label{action} 
 \end{equation}
where $M$ is a 5D fundamental scale and we assume that the 5D Dirac mass matrix $m$ is diagonal ($m_i\equiv 
m_{ii}$) for simplicity. The 5D gamma matrices $\Gamma^A$ are defined by
 \begin{equation}
  \Gamma^\mu = \begin{pmatrix} 0 & \sigma^{\mu} \\ \bar{\sigma}^{\mu} & 0 \end{pmatrix}~~~\mbox{ and }~~~\Gamma^5 =-i \begin{pmatrix} 1 & 0 \\ 0 & -1 \end{pmatrix}.
 \end{equation}
Note that the mass dimension of $\Psi_i(x,y)$ is $3/2$. The zero modes of 
$\Psi_i(x,y)$ follow Dirac equations as
 \begin{equation}
  (i\Gamma^5\partial_5+m_i)\Psi_i^{(0)}(x,y)=0.
 \end{equation}  
The solutions of the 5D Dirac equations determine the wave function profiles of the 
zero modes on the bulk as $e^{\mp m_iy}$ for $\chi_i(x,y)$ and 
$\overline{\psi_i}(x,y)$. In order to obtain chiral fermions in 4D, the fields 
are transformed under an orbifold parity as
 \begin{eqnarray}
  Z_2:~\Psi_i(x,y)\rightarrow P\Psi_i(x,y)=+\Psi_i(x,y),
 \end{eqnarray}
where $P = -i \Gamma_5$. Under this orbifold parity, only $\overline{\psi_i}(x,y)$ can have
 a zero mode. Note that the bulk masses $m_i$ have to carry negative $Z_2$ 
parity to make the fundamental action invariant under the above parity transformation. After canonical normalization of the Dirac fermions in 4D, the zero modes of $\Psi_{R_i}(x,y)\equiv(0,\overline{\psi_i}(x,y))^T$ are represented by a normalized field $\psi_{R_i}^{(0)}(x)$ as
 \begin{equation}
  \Psi_{R_i}^{(0)}(x,y)=\sqrt{\frac{2m_i}{(e^{2m_i\ell}-1)M}}
                        e^{m_iy} \psi_{R_i}^{(0)}(x), \label{5Dto4DSpinor}
 \end{equation}
where $\psi_{R_i}^{(0)}(x)$ are identified with the right-handed neutrinos in 4D, $\psi_{R_i}^{(0)}=\nu_{R_i}$. A point to note is that the 5D wave function profiles of $\Psi_{R_i}^{(0)}(x,y)$ with real positive (negative) $m_i$ localize at the hidden (SM) brane since the profile is described by the exponential function, $e^{m_iy}$. Therefore, the right-handed neutrinos, $\psi_{R_i}^{(0)}(=\nu_{R_i})$, have exponentially suppressed Yukawa couplings at the SM brane when $m_i\ell\gg1$ with real positive $m_i$.

The split seesaw mechanism utilizes the above property of the wave function 
profiles of the right-handed neutrinos. The relevant Lagrangian for the split 
seesaw mechanism is given by
 \begin{eqnarray}
  S &=& \int d^4x\int_0^\ell dy\Bigg[M
        \left(i\overline{\Psi^{(0)}_{R_i}}\Gamma^A\partial_A\Psi_{R_i}^{(0)} 
              +m_{i}\overline{\Psi_{R_i}^{(0)}}\Psi_{R_i}^{(0)}\right)
        \nonumber\\ 
    & & \phantom{\int d^4x\int_0^\ell dy\Big\{}\left.
        -\delta(y)\left((\tilde{y}_\nu)_{i\alpha}\overline{\Psi^{(0)}_{R_i}}
                        L_\alpha\phi
                        +\frac{(\tilde{M}_R)_{ij}}{2}
                         \overline{\Psi^{(0)c}_{R_i}}\Psi^{(0)}_{R_j} 
                        +{\rm h.c.}\right)\right]. \label{Sneutrino}
 \end{eqnarray}
Inserting \eqref{5Dto4DSpinor} into \eqref{Sneutrino}, we can obtain the effective 
4D Majorana mass matrix of the right-handed neutrinos and the neutrino Yukawa 
coupling matrix as
 \begin{eqnarray}
  (M_R)_{ij} &=& f_if_j(\tilde{M}_R)_{ij}, \label{MR4D} \\
  (y_\nu)_{i\alpha} &=& f_i(\tilde{y}_\nu)_{i\alpha}, \label{Yukawa4D}
 \end{eqnarray}
respectively, where 
 \begin{eqnarray}
  f_i\equiv\sqrt{\frac{2 m_i}{(e^{2 m_i \ell}-1)M}}.
 \end{eqnarray}
It is convenient for the following discussion to define a diagonal matrix 
$F$ as $F_{ij}\equiv\delta_{ij}f_j$. The effective 4D Majorana mass matrix 
of the right-handed neutrinos and the neutrino Yukawa coupling matrix given in 
\eqref{MR4D} and \eqref{Yukawa4D} are rewritten with the use of $F$,
 \begin{eqnarray}
  M_{R} &=& F\tilde{M}_{R}F, \label{MR4Dnew} \\
  y_\nu &=& F\tilde{y}_\nu. \label{Yukawa4Dnew}
 \end{eqnarray}
The seesaw mechanism leads to
 \begin{eqnarray}
  M_{\nu}
  =y_\nu^{T}M_R^{-1}y_\nu v^2=\tilde{y}_\nu^T\tilde{M}_R^{-1}\tilde{y}_\nu v^2. 
  \label{lightMajoranaMass}
 \end{eqnarray}
for the mass matrix of the light neutrinos. Note that the factors $f_i$ are completely canceled out in the seesaw mechanism. This is one of the interesting features of the split seesaw mechanism. 

The most important feature of the mechanism is as follows: By taking appropriate values for $m_i$, we can obtain strongly hierarchical mass spectra of the right-handed neutrinos and the neutrino Yukawa matrix elements shown in \eqref{MR4D} and \eqref{Yukawa4D} due to the exponential suppression in $f_i$, without introducing strongly hierarchical many-mass scales in the model. For instance, if one takes $(m_1\ell,m_2\ell,m_3\ell)\simeq(23.3,3.64,2.26)$, $M=5\times10^{17}$ GeV, $\ell^{-1}=10^{16}$ GeV, $(\tilde{M}_R)_{ii}=10^{15}$ GeV in a diagonal basis of $\tilde{M}_R$, we can obtain a splitting mass spectrum of the right-handed neutrinos including both keV and intermediate mass scales as
 \begin{eqnarray}
  (M_1,M_2,M_3)=(5\mbox{ keV},10^{11}\mbox{ GeV},10^{12}\mbox{ GeV}),
  \label{massspectrum}
 \end{eqnarray}
in the 4D effective theory. In this case, mass scales in the fundamental action are super 
heavy as $\mathcal{O}(10^{15-17})$ GeV or on the EW scale 
$\mathcal{O}(10^2)$ GeV, and the tiny active neutrino mass scale can be 
realized by the seesaw mechanism without tiny neutrino Yukawa couplings. 
Further, the lightest sterile neutrino with a keV mass could be a candidate for 
the DM and heavier sterile neutrinos with masses of 
$\mathcal{O}(10^{11-12})$ GeV could lead to the BAU via leptogenesis. 

\section{Localizations of right-handed neutrinos and Majorana masses}

\subsection{Localized right-handed neutrinos at the hidden brane and Majorana masses at the SM brane}

In the previous section, we have presented a brief review of the split seesaw 
mechanism. In this mechanism, the 5D wave functions of the right-handed 
neutrinos exponentially localize at the hidden brane \eqref{5Dto4DSpinor}, and 
the Majorana masses of the right-handed neutrinos are localized at the SM brane
 \eqref{Sneutrino}. As a result, one can easily obtain strongly hierarchical mass spectra of the right-handed neutrinos and the neutrino 
Yukawa couplings without introducing strongly hierarchical many-mass scales in the fundamental action. Regarding the 
resultant active neutrino mass after the seesaw mechanism, it is given by the 
same formula as in the type-I seesaw mechanism. The setup and results are 
summarized as
 \begin{eqnarray}
  \left\{
  \begin{array}{ll}
   \Psi_{R_i}^{(0)}(x,y)=f_ie^{m_iy}\psi_{R_i}^{(0)}(x)                                       & \mbox{ for the localization of the 5D wave functions} \\
   S\supset-\delta(y)\frac{(\tilde{M}_R)_{ij}}{2}\overline{\Psi^{(0)c}_{R_i}}\Psi^{(0)}_{R_j} & \mbox{ for the localization of the Majorana mass matrix}
  \end{array}
  \right., \label{l1}
 \end{eqnarray}
and 
 \begin{eqnarray}
  \left\{
  \begin{array}{ll}
   (M_R)_{ij}=f_if_j(\tilde{M}_R)_{ij}             & \mbox{ for the 4D Majorana mass matrix}   \\
   (y_\nu)_{i\alpha}=f_i(\tilde{y}_\nu)_{i\alpha}              & \mbox{ for the 4D neutrino Yukawa coupling matrix} \\
   M_{\nu}=\tilde{y}_\nu^T\tilde{M}_R^{-1}\tilde{y}_\nu v^2 & \mbox{ for the light (active) neutrino mass matrix} 
  \end{array}
  \right., \label{r1}
 \end{eqnarray}
with the suppression factors $f_i$ (if $m_i\ell\gg1$), respectively. We call 
this case $(\Psi_{R}^{(0)},\tilde{M}_R)=(H,S)$ where ``$H$'' and ``$S$'' mean
 that the 5D wave function ($\Psi_R^{(0)}$) or Majorana mass ($\tilde{M}_R$) of 
the right-handed neutrinos localize at the hidden and SM branes, respectively. 
Therefore, note that the other three possibilities of localizations 
($(\Psi_{R}^{(0)},\tilde{M}_R)=(S,H)$, $(H,H)$, and $(S,S)$) can be generically 
considered. In the following sections, we investigate the other three 
localizations and their results. 

\subsection{Localized right-handed neutrinos at the SM brane and Majorana masses at the hidden brane}

The split seesaw mechanism assumes $(H,S)$ localizations, namely the 5D 
wave functions of the right-handed neutrinos localize at the hidden brane while
 the Majorana mass matrix localizes at the SM brane. Here we consider the reverse 
situation as $(\Psi_{R}^{(0)},\tilde{M}_R)=(S,H)$, i.e. the 5D wave functions 
of the right-handed neutrinos localize at the SM brane while the Majorana 
mass matrix localizes at the hidden brane. The localization of the 5D 
wave function of the right-handed neutrinos is determined by the sign for the 
bulk mass in \eqref{action}. Therefore, if one take a minus sign for the bulk mass
 unlike the case of the split seesaw mechanism $(H,S)$, one can obtain a 5D 
wave function profile described by $e^{-m_iy}$ for the zero mode of the 
right-handed neutrinos. In this case, after normalization, the 
$\Psi_{R_i}^{(0)}(x,y)$ is written as 
 \begin{eqnarray}
  \Psi_{R_i}^{(0)}(x,y)=\sqrt{\frac{2m_i}{(1-e^{-2m_i\ell})M}}e^{-m_iy}
                        \psi_{R_i}^{(0)}(x)
                       \equiv g_ie^{-m_iy}\psi_{R_i}^{(0)}(x). \label{s}
 \end{eqnarray}
Note that the factor $g_i$ is not an exponential suppression one if $m_i\ell\gg1$, 
unlike the $f_i$ in the split seesaw mechanism. Therefore, the right-handed 
neutrinos couple to the SM particles without any exponential suppressions.

For the localization of the Majorana mass matrix at the hidden brane, the 
relevant action becomes
 \begin{eqnarray}
  S &=& \int d^4x\int_0^\ell dy\Bigg[M
        \left(i\overline{\Psi^{(0)}_{R_i}}\Gamma^A\partial_A\Psi_{R_i}^{(0)} 
              -m_i\overline{\Psi_{R_i}^{(0)}}\Psi_{iR}^{(0)}\right)
        \nonumber\\ 
    & & \phantom{\int d^4x\int_0^\ell dy\Big\{}\left.
        -\left(\delta(y)(\tilde{y}_\nu)_{i\alpha}\overline{\Psi^{(0)}_{R_i}}
               L_\alpha\phi
               +\delta(y-\ell)\frac{(\tilde{M}_R)_{ij}}{2}
                \overline{\Psi^{(0)c}_{R_i}}\Psi^{(0)}_{Rj}+{\rm h.c.}\right)
        \right]. \label{S2}
 \end{eqnarray}
After substituting \eqref{s} into \eqref{S2}, one can obtain the 4D Majorana 
mass matrix of the right-handed neutrinos and the 4D neutrino Yukawa coupling 
matrix as
 \begin{eqnarray}
  (M_R)_{ij}=g_ig_je^{-(m_i+m_j)\ell}(\tilde{M}_R)_{ij}
            =f_if_j(\tilde{M}_R)_{ij}, \qquad
  (y_\nu)_{i\alpha}=g_i(\tilde{y}_\nu)_{i\alpha},
 \end{eqnarray}
respectively, where the relation $f_i=g_ie^{-m_i\ell}$ is utilized. By taking the 
matrix forms as \eqref{MR4D} and $y_\nu=G\tilde{y}_\nu$ with 
$G_{ij}\equiv\delta_{ij}g_i$, the seesaw mechanism leads to 
$M_{\nu}=\tilde{y}_\nu^TE\tilde{M}_R^{-1}E\tilde{y}_\nu v^2$ where 
$E\equiv GF^{-1}=F^{-1}G$ and $E_{ij}=\delta_{ij}e^{m_i\ell}$.

We find that the 4D Majorana masses of the right-handed neutrinos are 
exponentially suppressed from the fundamental scale $\tilde{M}_R$, like the split 
seesaw mechanism. Therefore, strongly hierarchical mass spectra of the 
right-handed neutrinos can also be realized without introducing strongly 
hierarchical many-mass scales for the right-handed neutrinos. On the other hand, the 
neutrino Yukawa couplings are not suppressed, unlike the split seesaw case. As a result, the fundamental mass scale of the active neutrino mass matrix is enhanced
 compared with that in the type-I and split seesaw cases. These are summarized as
 \begin{eqnarray}
  \left\{
  \begin{array}{ll}
   \Psi_{R_i}^{(0)}(x,y)=g_ie^{-m_iy}\psi_{R_i}^{(0)}(x)                                       & \mbox{ for the localization of the 5D wave functions} \\
   S\supset-\delta(y-\ell)\frac{(\tilde{M}_R)_{ij}}{2}\overline{\Psi^{(0)c}_{R_i}}\Psi^{(0)}_{R_j} & \mbox{ for the localization of the Majorana mass matrix}
  \end{array}
  \right.,
 \end{eqnarray}
and 
 \begin{eqnarray}
  \left\{
  \begin{array}{ll}
   (M_R)_{ij}=f_if_j(\tilde{M}_R)_{ij}               & \mbox{ for the 4D Majorana mass matrix}   \\
   (y_\nu)_{i\alpha}=g_i(\tilde{y}_\nu)_{i\alpha}                & \mbox{ for the 4D neutrino Yukawa coupling matrix} \\
   M_{\nu}=\tilde{y}_\nu^TE\tilde{M}_R^{-1}E\tilde{y}_\nu v^2 & \mbox{ for the light (active) neutrino mass matrix} 
  \end{array}
  \right..
 \end{eqnarray}

\subsection{Localized right-handed neutrinos and Majorana masses at the hidden brane}

Next, we investigate the $(H,H)$ case, i.e., both the wave functions and the Majorana 
masses of the right-handed neutrinos are localized at the hidden brane. The 5D 
wave function profiles of the right-handed neutrinos are given by 
\eqref{5Dto4DSpinor}. Then the relevant action is described by
 \begin{eqnarray}
  S &=& \int d^4x\int_0^\ell dy\Bigg[M
        \left(i\overline{\Psi^{(0)}_{R_i}}\Gamma^A\partial_A\Psi_{R_i}^{(0)} 
              +m_i\overline{\Psi_{R_i}^{(0)}}\Psi_{iR}^{(0)}\right)
        \nonumber\\ 
    & & \phantom{\int d^4x\int_0^\ell dy\Big\{}\left.
        -\left(\delta(y)(\tilde{y}_\nu)_{i\alpha}\overline{\Psi^{(0)}_{R_i}}
               L_\alpha\phi
               +\delta(y-\ell)\frac{(\tilde{M}_R)_{ij}}{2}
                \overline{\Psi^{(0)c}_{R_i}}\Psi^{(0)}_{Rj}+{\rm h.c.}\right)
        \right]. \label{S3}
 \end{eqnarray}
Inserting \eqref{5Dto4DSpinor} into \eqref{S3}, we can find 
\begin{eqnarray}
  \left\{
  \begin{array}{ll}
   \Psi_{R_i}^{(0)}(x,y)=f_ie^{m_iy}\psi_{R_i}^{(0)}(x)                                       & \mbox{ for the localization of the 5D wave functions} \\
   S\supset-\delta(y-\ell)\frac{(\tilde{M}_R)_{ij}}{2}\overline{\Psi^{(0)c}_{R_i}}\Psi^{(0)}_{R_j} & \mbox{ for the localization of the Majorana mass matrix}
  \end{array}
  \right.,
 \end{eqnarray}
as the setup, and 
 \begin{eqnarray}
  \left\{
  \begin{array}{ll}
   (M_R)_{ij}=g_ig_j(\tilde{M}_R)_{ij}                         & \mbox{ for the 4D Majorana mass matrix}   \\
   (y_\nu)_{i\alpha}=f_i(\tilde{y}_\nu)_{i\alpha}                          & \mbox{ for the 4D neutrino Yukawa coupling matrix} \\
   M_{\nu}=\tilde{y}_\nu^TE^{-1}\tilde{M}_R^{-1}E^{-1}\tilde{y}_\nu v^2 & \mbox{ for the light (active) neutrino mass matrix} 
  \end{array}
  \right.,
 \end{eqnarray}
as the results. It is seen that the 4D Majorana masses do not include 
exponential suppression factors, and thus this $(H,H)$ case requires some mass 
scales for the right-handed neutrino sector in order to realize strongly 
hierarchical mass spectra. For the 4D neutrino Yukawa coupling, it is 
suppressed by $f_i$ as well as the split seesaw case $(H,S)$. The resultant 
active neutrino mass matrix includes a suppression factor $E^{-1}$ 
($(E^{-1})_{ij}=\delta_{ij}e^{-m_i\ell}$) compared to the type-I and split 
seesaw mechanisms.

\subsection{Localized right-handed neutrinos and Majorana masses at the SM brane}

Finally, we show the $(S,S)$ case, i.e., both the wave functions and the Majorana masses 
of the right-handed neutrinos are localized at the SM brane. The 5D 
wave function profiles of the right-handed neutrinos are given by \eqref{s} and 
the relevant action is described by
 \begin{eqnarray}
  S &=& \int d^4x\int_0^\ell dy\Bigg[M
        \left(i\overline{\Psi^{(0)}_{R_i}}\Gamma^A\partial_A\Psi_{R_i}^{(0)} 
              -m_{i}\overline{\Psi_{R_i}^{(0)}}\Psi_{R_i}^{(0)}\right)
        \nonumber\\ 
    & & \phantom{\int d^4x\int_0^\ell dy\Big\{}\left.
        -\delta(y)\left((\tilde{y}_\nu)_{i\alpha}\overline{\Psi^{(0)}_{R_i}}
                        L_\alpha\phi
                        +\frac{(\tilde{M}_R)_{ij}}{2}
                         \overline{\Psi^{(0)c}_{R_i}}\Psi^{(0)}_{R_j} 
                        +{\rm h.c.}\right)\right]. \label{S4}
 \end{eqnarray}
Substituting \eqref{s} into \eqref{S4}, we can find
\begin{eqnarray}
  \left\{
  \begin{array}{ll}
   \Psi_{R_i}^{(0)}(x,y)=g_ie^{-m_iy}\psi_{R_i}^{(0)}(x)                                       & \mbox{ for the localization of the 5D wave functions} \\
   S\supset-\delta(y)\frac{(\tilde{M}_R)_{ij}}{2}\overline{\Psi^{(0)c}_{R_i}}\Psi^{(0)}_{R_j} & \mbox{ for the localization of the Majorana mass matrix}
  \end{array}
  \right.,
 \end{eqnarray}
as the setup, and 
 \begin{eqnarray}
  \left\{
  \begin{array}{ll}
   (M_R)_{ij}=g_ig_j(\tilde{M}_R)_{ij}                         & \mbox{ for the 4D Majorana mass matrix}   \\
   (y_\nu)_{i\alpha}=g_i(\tilde{y}_\nu)_{i\alpha}                          & \mbox{ for the 4D Yukawa coupling matrix} \\
   M_{\nu}=\tilde{y}_\nu^T\tilde{M}_R^{-1}\tilde{y}_\nu v^2 & \mbox{ for the light (active) neutrino mass matrix} 
  \end{array}
  \right.,
 \end{eqnarray}
as the results. It can be seen that both the 4D Majorana masses of the right-handed
 neutrinos and the neutrino Yukawa couplings are not suppressed because of the 
factor $g_i$ (if $m_i\ell\gg1$), and the active neutrino mass matrix is 
described by the same formula as in the type-I and split seesaw mechanisms.
 Therefore, this $(S,S)$ case is similar to the type-I seesaw because of the 
localizations of the wave functions and Majorana masses at the SM brane. 

\section{Separate seesaw}

We propose a new seesaw model using of the above results. A fundamental 
assumption in this seesaw model is that each generation of the right-handed 
neutrinos can have different localizations of the 5D wave function and Majorana 
mass, e.g., one of the generations of the right-handed neutrinos has $(H,S)$ 
localization and the others have $(S,H)$ localization. This means that the 
localizations of $(\Psi_{R}^{(0)},\tilde{M}_R)$ are separate for each 
generation of right-handed neutrinos. Under this assumption, when one 
introduces two generations of right-handed neutrinos into the SM, one can 
consider 10 possible combinations of the localizations,\footnote{Here we do not 
consider the ordering of generations; for instance, we do not distinguish the $((H,S),(S,H))$ case from the $((S,H),(H,S))$ one. There are 24 possible 
combinations of localizations in the case of three generations of right-handed neutrinos.} e.g., 
$((\Psi_{R_1}^{(0)},\tilde{M}_{R_1}),(\Psi_{R_2}^{(0)},\tilde{M}_{R_2}))=((H,S),(H,S)),$ $((H,S),(S,H)),$ $((H,S),(H,H)),$ $((H,S),(S,S)),$ and 
$((S,H),(S,H))$ etc., where $\tilde{M}_{R_i}$ is the fundamental mass scale of the $i$th generation of the right-handed neutrinos in a 5D action with a diagonal 
basis for the 5D Dirac mass matrix. Therefore, in this seesaw model, $\tilde{M}_R$ 
is not generically matrix. In this work, we investigate a combination and its 
results as an example, which will be interestingly applied to DM and baryogenesis models with respect to a left-right mixing angle as we will see later.

We consider the case of 
$((\Psi_{R_1}^{(0)},\tilde{M}_{R_1}),(\Psi_{R_2}^{(0)},\tilde{M}_{R_2}))=((H,S),(S,H))$ 
as an example with two generations of right-handed neutrinos for 
simplicity.\footnote{Such a separate localization of the Majorana masses can be 
realized by introducing gauge singlet scalars with lepton number and an 
additional discrete symmetry. We will show an example of a realistic case of 
three generations in the next section.} The relevant action for this case of the {\it 
separate} seesaw is given by
 \begin{eqnarray}
  S &=& \int d^4x\int_0^\ell dy\Bigg[M
        \left\{i\overline{\Psi^{(0)}_{R_i}}\Gamma^A\partial_A\Psi_{R_i}^{(0)} 
               +\left(m_1\overline{\Psi_{R_1}^{(0)}}\Psi_{R_1}^{(0)}
                      -m_2\overline{\Psi_{R_2}^{(0)}}\Psi_{R_2}^{(0)}
                \right)\right\} \nonumber\\ 
    & & \left.
        -\left\{\delta(y)\left((\tilde{y}_\nu)_{i\alpha}\overline{\Psi^{(0)}_{R_i}}
                               L_\alpha\phi
                               +\frac{\tilde{M}_1}{2}
                                \overline{\Psi^{(0)c}_{R_1}}\Psi^{(0)}_{R_1}
                         \right)
                +\delta(y-\ell)\frac{\tilde{M}_2}{2}
                 \overline{\Psi^{(0)c}_{R_2}}\Psi^{(0)}_{R_2}+{\rm h.c.}\right\}
        \right],
 \end{eqnarray}
where $i=1,2$ and $\tilde{M}_i$ are the fundamental mass scales of each generation 
of right-handed neutrinos. Note that the Majorana masses of the right-handed 
neutrinos ($\tilde{M}_i$) in this action are not described by a matrix form 
because of the separate localizations at different branes. The solutions of the 5D 
Dirac equations from this action determine the extra-dimensional wave function 
profiles of the right-handed neutrinos as
 \begin{eqnarray}
  \Psi_{R_1}^{(0)}(x,y)=f_1e^{m_1y}\psi_{R_1}^{(0)}(x),\qquad
  \Psi_{R_2}^{(0)}(x,y)=g_2e^{-m_2y}\psi_{R_2}^{(0)}(x). 
 \end{eqnarray}
Then the resultant 4D Majorana masses, 4D Yukawa coupling matrices, and the 
light neutrino mass matrix after the separate seesaw are written down as
 \begin{align}
  &M_i=f_i^2\tilde{M}_i, \\
  &(y_\nu)_{1\alpha}=f_1(\tilde{y}_\nu)_{1\alpha},\qquad 
   (y_\nu)_{2\alpha}=g_2(\tilde{y}_\nu)_{2\alpha}, \\
  &(M_\nu)_{\alpha\beta}=((\tilde{y}_\nu)_{1\alpha}(\tilde{y}_\nu)_{1\beta}\tilde{M}_1^{-1}
                          +e^{2m_2\ell}(\tilde{y}_\nu)_{2\alpha}(\tilde{y}_\nu)_{2\beta}
                           \tilde{M}_2^{-1})v^2,
 \end{align}
where $\alpha,\beta=e,\mu,\tau$. After diagonalizing the light neutrino mass 
matrix $M_\nu$, one can obtain three mass eigenvalues of the light neutrinos 
as
 \begin{eqnarray}
  m_{\nu1}=0,\qquad 
  m_{\nu2}\simeq\mathcal{O}(|(\tilde{y}_\nu)_{1\alpha}|^2v^2\tilde{M}_R^{-1}),
  \qquad 
  m_{\nu3}\simeq\mathcal{O}(e^{2m_2\ell}|(\tilde{y}_\nu)_{2\alpha}|^2v^2\tilde{M}_R^{-1}),
 \end{eqnarray}
where we assume $|(\tilde{y}_\nu)_{i\alpha}|\simeq|(\tilde{y}_\nu)_{i\beta}|$ 
($\alpha\neq\beta$) and take $\tilde{M}_R$, which is a fundamental mass scale of
 the right-handed neutrino sector in the action, as 
$\tilde{M}_R\simeq\tilde{M}_1\simeq\tilde{M}_2$. Note that this kind of separate
 seesaw can lead to hierarchical mass eigenvalues of the light neutrinos, which 
are determined by exponential factors, due to the different localizations of the 
wave function and Majorana mass for each generation of right-handed 
neutrinos, unlike the type-I and split seesaw mechanisms. This model can also 
realize hierarchical mass spectra of the right-handed neutrinos and the neutrino 
Yukawa couplings among each generation of right-handed neutrinos. Other 
combinations of localizations lead to different patterns of mass spectra for the 
light neutrinos and right-handed neutrinos, and neutrino Yukawa couplings 
for each generation. Such formalism for all cases and extensions to three 
generations of right-handed neutrino models are straightforward. We now 
concentrate on this type of separate seesaw, 
$((\Psi_{R_1}^{(0)},\tilde{M}_{R_1}),(\Psi_{R_2}^{(0)},\tilde{M}_{R_2}))=((H,S),(S,H))$, 
and apply it to realistic models of neutrino DM and baryogenesis with three 
generations of right-handed neutrinos in the next section.

\section{Applications to models of dark matter and baryogenesis}

In this section, we apply the separate seesaw to models of neutrino DM with baryogenesis. It is known that a sterile neutrino with a keV scale mass can
 be a candidate for DM. Further, models with three generations of right-handed neutrinos including the keV sterile neutrino can simultaneously explain
 the BAU, the tiny active neutrino mass scale, and the LSND/MiniBooNE anomaly in 
addition to the 
DM~\cite{Asaka:2005an,Kusenko:2010ik,Adulpravitchai:2011rq,BRZ}. In
 such models, the tiny active neutrino mass is described by the type-I 
seesaw formula, and the models can be embedded into the split seesaw mechanism 
in order to realize split mass spectra without introducing strongly hierarchical
many-mass scales. However, it is also known that, in such models with the keV sterile 
neutrino, the sterile neutrino (with keV mass) should not contribute to the 
active neutrino masses to satisfy cosmological ($X$-ray) bounds on left-right 
mixing (neutrino Yukawa couplings) for the keV sterile neutrino as 
 \begin{eqnarray}
  \theta^2\equiv\sum\frac{|(y_\nu)_{1\alpha}|^2v^2}{M_1^2}\lesssim\mathcal{O}(10^{-11}-10^{-9}),
 \end{eqnarray} 
in a mass region of the sterile neutrino of $\mathcal{O}(1)$ 
keV$\lesssim M_s(\equiv M_1)\lesssim\mathcal{O}(10)$ keV (e.g., 
see~\cite{Kusenko:2009up} and references therein). This cosmological bound means
 that the resultant neutrino mass, $m_{\nu1}$, from the keV sterile neutrino 
after the seesaw mechanism is much smaller than the solar and atmospheric neutrino 
mass scales as $m_{\nu1}\ll m_{\rm sol}<m_{\rm atm}$, where $m_{\rm 
sol}\simeq8.73\times10^{-3}$ eV and $m_{\rm atm}\simeq5.05\times10^{-2}$ 
eV~\cite{Tortola:2012te}. In other words, once the keV sterile neutrino 
contributes to the active neutrino mass through the seesaw mechanism, the model 
is in conflict with the cosmological bound because of the large left-right mixing 
angle, 
 \begin{equation}
  \theta^2=\sum\frac{|(y_\nu)_{1\alpha}|^2v^2}{M_1^2}\simeq\frac{m_{\rm sol 
(atm)}}{M_s}\simeq\mathcal{O}(10^{-6}-10^{-5}).
 \end{equation} 
Therefore, in a realistic model with such keV sterile neutrino DM, an additional
 suppression for $\theta$ (equivalent to $|(y_\nu)_{1\alpha}|$) is needed to ensure consistency with the cosmological constraint. Further, at least two more 
generations of right-handed neutrinos are required to reproduce the solar 
and atmospheric neutrino mass scales. Such a situation is realized by taking the 
Yukawa couplings of the keV sterile neutrino to be additionally small compared 
to those of the other two sterile neutrinos, which give the solar and atmospheric
 neutrino mass scales, in some models with keV sterile neutrino DM. The separate
 seesaw can lead to the additional suppression of the left-right mixing angle 
(the Yukawa couplings) and give a typical mass spectrum of active neutrinos as 
$m_{\nu1}\ll m_{\nu2}(\simeq m_{\rm sol})<m_{\nu3}(\simeq m_{\rm atm})$ in the 
keV sterile neutrino DM models.

In order to see this, we extend the separate seesaw with two generations of right-handed neutrinos given in the previous section to the three-generations case. The relevant Lagrangian is given by 
 \begin{eqnarray}
  S &=& \int d^4x\int_0^\ell dy\Bigg[M
        \left\{i\overline{\Psi^{(0)}_{R_i}}\Gamma^A\partial_A\Psi_{R_i}^{(0)} 
               +\left(m_1\overline{\Psi_{R_1}^{(0)}}\Psi_{R_1}^{(0)}
                      -m_j\overline{\Psi_{R_j}^{(0)}}\Psi_{R_j}^{(0)}
                \right)\right\} \nonumber\\ 
    & & \left.
        -\left\{\delta(y)\left((\tilde{y}_\nu)_{i\alpha}\overline{\Psi^{(0)}_{R_i}}
                               L_\alpha\phi
                               +\frac{\tilde{M}_1}{2}
                                \overline{\Psi^{(0)c}_{R_1}}\Psi^{(0)}_{R_1}
                         \right)
                +\delta(y-\ell)\frac{\tilde{M}_j}{2}
                 \overline{\Psi^{(0)c}_{R_j}}\Psi^{(0)}_{R_j}+{\rm h.c.}\right\}
        \right],
  \label{separate}
 \end{eqnarray}
 where $i=1,2,3$, $j=2,3$, and we take a $2\times2$ diagonal mass matrix, 
$\tilde{M}_R={\rm Diag}\{\tilde{M}_2,\tilde{M}_3\}$.\footnote{Such a separate 
localization of the Majorana masses can be realized by introducing gauge singlet
 scalars with an additional discrete symmetry. For instance, we introduce two 
gauge singlet scalars with lepton number, which are localized at the SM or the hidden 
brane, i.e., $\phi_S$ is localized at the SM brane while $\phi_H$
 is localized at the hidden brane. Further, we impose a $Z_6$ discrete symmetry 
and assign charges under $Z_6$ as $Q(\phi_S)=1$, $Q(\phi_H)=\omega_6^4$, 
$Q(\Phi_{R_1}^{(0)})=1$, $Q(\Phi_{R_2}^{(0)})=\omega_6$, and 
$Q(\Phi_{R_3}^{(0)})=\omega_6^4$, where $\omega_6\equiv e^{i\pi/3}$. The setup 
can induce Majorana masses in the second line of \eqref{separate} after the
 lepton number violation by the VEVs of $\phi_S$ and $\phi_H$ as 
$\delta(y)\tilde{M_1}=\delta(y)\lambda_1\langle\phi_S\rangle$ and 
$\delta(y-\ell)\tilde{M}_j=\delta(y-\ell)\lambda_j\langle\phi_H\rangle$, where 
$\lambda_i$ are the Yukawa couplings among the right-handed neutrinos and new 
gauge singlets. For the neutrino Yukawa interactions, we have to assign 
appropriate charges to $L_\alpha$ with $Q(\phi)=1$ without reducing the rank of 
the neutrino Dirac mass matrix in order to realize three mass eigenvalues of the
 active neutrinos. If one introduces more gauge singlets, which do not have the 
lepton number but have $Z_6$ charges, one can obtain rich flavor structures 
in the neutrino sector to realize experimentally observed values of lepton 
mixing angles. This is the one of the examples inducing the above separate 
localization of the right-handed Majorana masses.} Note that this is an 
extension of the $((\Psi_{R_1}^{(0)},\tilde{M}_{R_1}),(\Psi_{R_2}^{(0)},\tilde{M}_{R_2}))=((H,S),(S,H))$ case to the three-generation case, 
$((\Psi_{R_1}^{(0)},\tilde{M}_{R_1}),(\Psi_{R_j}^{(0)},\tilde{M}_{R_j}))=((H,S),(S,H))$. In a similar manner to the two-generation case, we obtain 
extra-dimensional wave function profiles of the right-handed neutrinos as
 \begin{eqnarray}
  \Psi_{R_1}^{(0)}(x,y)=f_1e^{m_1y}\psi_{R_1}^{(0)}(x),\qquad
  \Psi_{R_j}^{(0)}(x,y)=g_je^{-m_jy}\psi_{R_j}^{(0)}(x),
 \end{eqnarray}
and read the resultant 4D Majorana masses, 4D neutrino Yukawa coupling matrices, light neutrino mass matrix, and three mass eigenvalues of the light neutrinos as
 \begin{align}
  &M_i=f_i^2\tilde{M}_i, \\
  &(y_\nu)_{1\alpha}=f_1(\tilde{y}_\nu)_{1\alpha},\qquad 
   (y_\nu)_{j\alpha}=g_j(\tilde{y}_\nu)_{j\alpha}, \\
  &(M_\nu)_{\alpha\beta}=\left((\tilde{y}_\nu)_{1\alpha}(\tilde{y}_\nu)_{1\beta}\tilde{M}_1^{-1}
                          +\sum_{j=2,3}e^{2m_j\ell}(\tilde{y}_\nu)_{j\alpha}(\tilde{y}_\nu)_{j\beta}
                           \tilde{M}_j^{-1}\right)v^2, \\
  &m_{\nu1}\simeq\mathcal{O}(|(\tilde{y}_\nu)_{1\alpha}|^2v^2\tilde{M}_R^{-1}),
   \qquad 
   m_{\nu j}\simeq\mathcal{O}(e^{2m_j\ell}|(\tilde{y}_\nu)_{j\alpha}|^2v^2\tilde{M}_R^{-1}),
 \end{align}
respectively. This separate seesaw can realize strongly hierarchical mass spectra of the sterile neutrinos and neutrino Yukawa couplings among the first and other generations of right-handed neutrinos. As a result, one of the three neutrino mass eigenvalues $m_{\nu1}$ is suppressed from the other two mass eigenvalues 
$m_{\nu j}$. In a realistic model, these two mass eigenvalues are of the order of the solar
 and atmospheric scales, $m_{\nu j}\sim m_{\rm sol(atm)}$. However, in the 
expressions of the mass eigenvalues $m_{\nu j}$, an exponential factor 
$e^{2m_j\ell}$ appears, while $m_{\nu1}$ does not include the factor. This 
affects the evaluation of left-right mixing as
 \begin{eqnarray}
  \theta^2=\sum\frac{|(y_\nu)_{1\alpha}|^2v^2}{M_1^2}\simeq\frac{m_{\rm sol 
(atm)}}{e^{2m_j\ell}M_s}, \label{ad}
 \end{eqnarray}
where we assume that all $|(\tilde{y}_\nu)_{i\alpha}|$ are of the same order.
 
We apply this separate seesaw to two typical mass spectra of the sterile neutrinos, which can give the sterile neutrino DM and generate 
the BAU in addition to the tiny active neutrino mass scales. The first example of
 the mass spectra is
 \begin{equation}
  (M_1,M_2,M_3)\sim
  (\mathcal{O}(\mbox{keV}),\mathcal{O}(10^{11}\mbox{ GeV})
   ,\mathcal{O}(10^{12}\mbox{ GeV})). \label{ex1}
 \end{equation}
 In this type of model, the lightest sterile neutrino with keV mass is the 
DM, and heavy sterile neutrinos with an intermediate mass scale can lead to the 
BAU via leptogenesis, as mentioned in the split seesaw mechanism. This mass 
spectrum of the sterile neutrinos requires corresponding neutrino Yukawa 
couplings $|(y_\nu)_{j\alpha}|$ of order 
$|(y_\nu)_{j\alpha}|\simeq\sqrt{m_{\rm 
sol(atm)}M_j}/v\sim\mathcal{O}(10^{-2}-10^{-1})$, as known in the canonical type-I seesaw mechanism. In our realization using the separate seesaw, when we take 
$(m_1\ell,m_2\ell,m_3\ell)\simeq(23.3,3.64,2.26)$, $M=5\times10^{17}$ GeV, 
$\ell^{-1}=10^{16}$ GeV, $\tilde{M}_R=\tilde{M}_i=10^{15}$ GeV, and 
$|(\tilde{y}_\nu)_{i\alpha}|\simeq(5-10)\times10^{-2}$ for all Yukawa couplings, the model leads to a mass 
spectrum \eqref{ex1}, and the solar and atmospheric neutrino mass scales. 
Further, an additional suppression for left-right mixing angle can be 
naturally realized as shown in \eqref{ad}, $\theta^2\sim\mathcal{O}(10^{-10})$, 
which is consistent with the cosmological bound. In this realization, the mass 
scales in the model are super-heavy $\mathcal{O}(10^{15-17})$ GeV or on the EW 
scale.

The next application of the separate seesaw is to a model with a mass 
spectrum of the sterile neutrinos as
 \begin{equation}
  (M_1,M_2,M_3)\sim(\mathcal{O}(\mbox{keV}),\mathcal{O}(10\mbox{ GeV}),\mathcal{O}(10\mbox{ GeV})). \label{ex2}
 \end{equation}
This type of model can explain the BAU through the mechanism proposed 
in~\cite{Akhmedov:1998qx}, i.e., an oscillation between the second and third generations of right-handed neutrinos. The model described by the mass 
spectrum \eqref{ex2} is known as the $\nu$MSM~\cite{Asaka:2005an}, 
which can also realize the tiny active neutrino masses through the seesaw 
mechanism by utilizing two heavier sterile neutrinos. Since this model also 
includes the keV sterile neutrino DM, the model should require an additional 
suppression for the left-right mixing (neutrino Yukawa couplings) of the keV 
sterile neutrino to satisfy the cosmological bound as well as the previous example. 
In the realization of the separate seesaw, when we take 
$(m_1\ell,m_2\ell,m_3\ell)\simeq(23.3,15.9,15.9)$, $M=5\times10^{17}$ GeV, 
$\ell^{-1}=10^{16}$ GeV, $\tilde{M}_R=\tilde{M}_i=10^{15}$ GeV, and 
$|\tilde{y}_{j\alpha}|\simeq\mathcal{O}(10^{-7})$, the model can lead to a 
mass spectrum \eqref{ex2} and $\theta^2<\mathcal{O}(10^{-11})$ in addition to both the solar and atmospheric neutrino
 mass scales.\footnote{We have taken smaller values of neutrino Yukawa couplings
 in the action as $|\tilde{y}_{j\alpha}|\simeq\mathcal{O}(10^{-7})$, which are 
the same as those required in the $\nu$MSM for the Yukawa couplings of the second and third generations of right-handed neutrinos. The split seesaw does not need such 
tiny Yukawa couplings in the action for the heavier right-handed neutrinos but 
this separate seesaw does require them. Therefore, an additional mechanism to 
give such tiny Yukawa couplings, e.g., the Froggatt-Nielsen 
mechanism~\cite{Froggatt:1978nt}, should be introduced.} 

\section{Summary}

The introduction of right-handed neutrinos into the SM can solve several problems in particle physics and cosmology, such as the DM, the generation of the BAU, the neutrino anomaly, and the realization of the small active neutrino mass through the seesaw mechanism. Further, extra-dimensional theories can also lead to various interesting results for phenomenology. In this work, we have focused on the case in which only right-handed neutrinos live in flat 5D space. In particular, the localizations of the extra-dimensional wave functions and Majorana masses of the right-handed neutrinos, and the results for relevant parameters for the neutrino masses have been investigated. First, it has been shown that different localizations of the wave functions and Majorana masses of the right-handed neutrinos in the extra dimension lead to different suppression factor dependences of the effective Majorana neutrino masses and the neutrino Yukawa couplings. 

Next, a new extra-dimensional seesaw model has been proposed by combining the above results. A fundamental assumption of the new seesaw is that each generation of right-handed neutrinos has different localizations for the wave function and Majorana masses in the extra dimension. As a result, the Majorana neutrino masses and the neutrino Yukawa couplings for each generation can be different to those of the canonical type-I seesaw and split seesaw mechanisms. We have shown the results of the case, $((\Psi_{R_1}^{(0)},\tilde{M}_{R_1}),(\Psi_{R_2}^{(0)},\tilde{M}_{R_2}))=((H,S),(S,H))$, as an example of the case of two generations of right-handed neutrinos. It has been found that this separate seesaw model case can lead to hierarchical mass eigenvalues of the light neutrinos, which are determined by exponential factors, due to the different localizations of wave function and Majorana masses for each generation of the right-handed neutrinos, unlike the type-I and split seesaw mechanisms. Further, it has also been shown that this model can realize hierarchical mass spectra of right-handed neutrinos and the neutrino Yukawa couplings among each generation of the right-handed neutrinos.

Finally, we have applied the results of the separate seesaw to models of the keV sterile neutrino DM with baryogenesis. It is known that an additional suppression for the left-right mixing angle of the keV sterile neutrino is needed to be consistent with the cosmological constraint in the framework of the seesaw mechanism. We have shown that the above separate seesaw extended to three generations of right-handed neutrinos can naturally realize the additional suppression for the mixing and be favored in two typical models of the keV sterile neutrino DM with baryogenesis.

At the end of the paper, we comment on realizations of leptonic mixing angles determined by neutrino oscillation experiments. The separate seesaw can be consistent with two possible discussions to obtain experimentally observed values of leptonic mixing angles, i.e., a discussion of $A_4$ flavor models in the split seesaw mechanism~\cite{Adulpravitchai:2011rq} and a minimal neutrino texture analysis for the model with the keV sterile neutrino DM~\cite{Shimizu:2012ry}. The fundamental assumption of the separate seesaw might also be interesting for the other phenomenological applications and/or other types of seesaw mechanism, such as a radiative seesaw model~\cite{Ma:2006km}. These applications will be presented in a different publication~\cite{AT}.
 
\subsection*{Acknowledgments}
This work is supported by Research Fellowships of the Japan Society for
 the Promotion of Science for Young Scientists. The author thanks M. Aoki, T. Asaka, N. Haba, H. Nakano, and K. Yoshioka for fruitful discussions at early stages of this work.


\end{document}